\documentclass{article}
\usepackage[utf8]{inputenc}
\usepackage{authblk}
\usepackage{hyperref}
\usepackage{xcolor}
\usepackage{graphicx}
\usepackage{makecell}
\usepackage[normalem]{ulem}
\usepackage{adjustbox}
\usepackage{amsmath}

\begin{document}

\title{Temporal Substepping Scheme for Magnetohydrodynamics with Cell-based Adaptive Mesh Refinement on Staggered Grid}
\author[1]{Ilja Honkonen}
\author[1]{Riku Jarvinen}
\author[1]{David Phillips}
\affil[1]{Finnish Meteorological Institute, Helsinki, Finland}
\date{\today}
\maketitle

\abstract
We present a new algorithm for numerical magnetohydrodynamics on staggered meshes preserving $\nabla \cdot B = 0$.
Our algorithm is based on the constrained transport method and supports both cell-based adaptive mesh refinement and temporal substepping.
We handle resolution changes directly on the logically Cartesian grid without needing interpolation or projection between nested or neighboring grids, nor coupling the solution between refinement levels.

\section{Introduction}
\label{sec:introduction}

Astrophysical plasmas are widely studied using numerical magnetohydrodynamics (MHD).
These include solar system plasmas such as the Sun, the solar wind, planetary and cometary magnetospheres, the heliosphere, and others \cite{flash00,gumics,pluto12,enzo14,amrvac18,athena20}.
Many established methods in computational fluid dynamics (CFD) form the foundation of finite-volume methods (FVM) used in MHD \cite{leveq2,toth00}.
In MHD however, electric and magnetic fields are also included which adds an extra layer of complexity compared to CFD.

One of the key approaches to satisfying the geometric constraints inherent in Maxwell’s equations of MHD and kinetic plasma models are staggered mesh-based solvers \cite{marti15,mignone21}.
This approach is particularly important for maintaining the divergence-free magnetic field condition
\begin{align*}
\nabla \cdot \vec{B} & = 0.
\end{align*}
The staggered grid was first introduced in CFD with volume-averaged, cell-centered pressure and area-averaged, face-centered velocity \cite{harlow65}.
Yee developed the staggered mesh technique, nowadays called the \emph{Yee lattice}, for the curl vector operators in Maxwell’s equations and advanced the system in time with a leapfrog scheme in electromagnetic simulations \cite{yee66}.
In this approach, components of electric and magnetic fields are discretized at offset spatial locations, and the technique became foundational in MHD and space plasma models in general.
Enforcing the $\nabla \cdot \vec{B} = 0$ condition was a key step in addition to the formulations obeying the conservation of mass, momentum and energy \cite{ryu95,powell99}.
This mitigated nonphysical forces, violations of conservation laws and different numerical instabilities.

A major next step in $\nabla \cdot \vec{B} = 0$ preserving algorithms was the introduction of line-average electric fields centered on cell edges for propagating magnetic fields on cell faces - termed the constrained transport (CT) method \cite{brecht81,evans88,balsara99}.
The CT scheme geometrically ensures the vanishing time derivative of the magnetic field divergence to machine precision in several dimensions
\begin{align*}
\frac{\partial}{\partial t}(\nabla \cdot \vec{B}) & = -\nabla \cdot (\nabla \times \vec{E}) = 0.
\end{align*}
Since then, many divergence-free algorithms have been developed for MHD and kinetic space plasma codes including, notably, Godunov-type methods with CT, enabling shock-capturing FVM that satisfy the $\nabla \cdot \vec{B} = 0$ condition \cite{balsara99,toth00}.
Further works extended these to higher-order reconstructions, weighted essentially non-oscillatory (WENO) schemes, adaptive spatial resolution, and divergence-free interpolation \cite{balsara01,balsara04,balsarakim04}.

Several other approaches to control $\nabla \cdot \vec{B}$ have also been developed such as hyperbolic cleaning, eight-wave and projection methods; see e.g.~\cite{toth00} and references therein.

Other foundational techniques in space plasma codes are e.g.~adaptive mesh refinement (AMR) and local time stepping or temporal substepping or subcycling.
They are widely applied for controlling spatial and temporal resolution in the simulation domain since the 1990s, see e.g.~\cite{lipatov02,gumics,toth12} and references therein.

Many approaches exists to solve each refinement level separately and then to apply divergence-preserving interpolation, extrapolation or projection methods for coupling together the solution at changes in spatial resolution.
For example, nested grids of uniform resolution are used in \cite{balsara01}, which are solved independently using non-AMR solvers.
An advantage of such couplings is that the different refinement levels can use their own timesteps in a straightforward way.
On the other hand, challenges include constantly translating quantities between nested grids while maintaining $\nabla \cdot B = 0$.

Here we present a new $\nabla \cdot B = 0$ preserving CT staggered mesh algorithm supporting both cell-by-cell AMR and temporal substepping.
Our algorithm handles resolution changes directly on the logically Cartesian grid and does not require e.g.~coupling of solutions, interpolation or projection between nested or neighboring grids.
In section \ref{sec:staggered} we describe the MHD flux-based staggered solver scheme our new algorithm is derived from \cite[BS99 hereafter]{balsara99}.
In section \ref{sec:substepping} we present the extension of the BS99 scheme to temporal substepping and in section \ref{sec:amr} we add cell-by-cell AMR.
We demonstrate our approach with several tests in section \ref{sec:tests}, and discuss the results in section \ref{sec:discussion} with conclusions in section \ref{sec:conclusions}.

\section{Magnetic field on staggered mesh}
\label{sec:staggered}

The BS99 approach uses face-centered fluxes of the  magnetic field, calculated by an FVM 1d MHD solver from cell-centered values, to advance a divergence-free solution of the face-centered - staggered - magnetic field in 3d.
The change to cell-centered magnetic field from the original 1d solution is ignored as the cell-centered field is derived from the face-centered field.
Optionally, total energy density is adjusted in order to account for different magnetic pressures between original and staggered solutions.

Another benefit of the BS99 approach is that ``extra'' features of the 1d solver can be used easily, such as the $B_0+B_1$ split described in \cite{tanaka94}.
As shown in \cite{toth00}, in a uniform mesh a staggered representation of magnetic field is not necessary for achieving divergence-free solution in 3d.
On the other hand, we show in section \ref{sec:amr} that the approach of BS99 is simple to extend to a cell-by-cell refined grid and does not require additional interpolation/prolongation/restriction operations as presented e.g.~in \cite{toth02}.

\begin{figure}
\centering
\includegraphics[width = \textwidth]{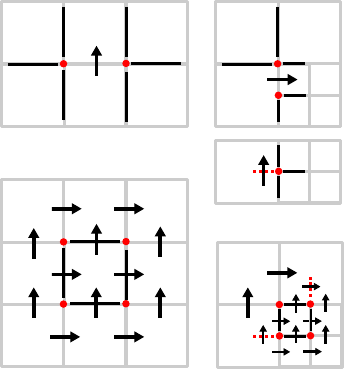}
\caption{
Illustration of relation between MHD flux calculations and changes to face magnetic fields in staggered field solver.
}
\label{fig:grid}
\end{figure}

In order to simplify the implementation of temporal substepping, we do not store the intermediate electric fields calculated from MHD fluxes, and instead we directly store the final change to magnetic field of adjacent faces.
Top left panel of figure \ref{fig:grid} illustrates which face magnetic fields (black bars) are affected by one flux calculation (arrow) in uniform mesh, with implicit edge electric fields represented by red circles.
Bottom left panel shows all fluxes required for advancing the middle cell's magnetic field during one timestep.
Not only are fluxes in/out of the cell itself required, but fluxes between all cells sharing at least one edge with middle cell must also be solved.

\section{Temporal substepping}
\label{sec:substepping}

In order to obtain a mathematically correct solution, the length of timestep in each simulation cell is limited by the fastest MHD wave speed propagating through it from its 6 faces \cite{cfl}.
A trivial approach is to limit the length of timestep of all simulation cells to the shortest step allowed in any cell of the simulation.
In case the length of allowed timestep differs by a large factor in different cells, using the shortest required timestep everywhere can increase both numerical diffusion as well as time to solution significantly due to having to solve some cells many more times than necessary.
We decrease this problem by implementing temporal substepping in which cells with a longer allowed timestep are solved less often than cells with a shorter allowed timestep.
The same approach is used e.g.~in the GUMICS model of Earth's magnetosphere, albeit with a cell-centered magnetic field and periodic cleaning of $\nabla \cdot \vec B$ \cite{gumics}.

For each simulation cell $i$ we calculate its longest allowed timestep $\Delta t_i$ and define the length of one simulation substep as the shortest of all allowed timesteps $\delta t = min(\Delta t_i)$.
We also define for each cell a substep level $L_i$ as the largest non-negative integer that fulfills $2^{L_i} \delta t \leq \Delta t_i$, and a maximum substep level for the simulation $L_{max} = max(L_i)$.
One timestep of the simulation therefore consists of $2^{L_{max}}$ subteps during which cell $i$ is solved $2^{L_i}$ times.

In order to preserve divergence-free condition of magnetic field we update the normal component of magnetic field ($B_N$) simultaneously at every face of one cell, along with plasma density, momentum and total energy (MHD state).
Two cells with different substep periods that share a face must therefore both track changes to $B_N$ of that face separately so that $\vec B$ in either cell can be updated independently and $\nabla \cdot \vec B = 0$ maintained.
Hence $B_N$ and its change are stored on all 6 faces of each cell and every face has two versions of $B_N$.
Any change to $B_N$ is recorded in both cells, regardless of their logical size in case of AMR.
At the end of every timestep, consisting of one or more substeps, we assign the average $B_N$ of a face to both cells in order to keep the solution consistent.
We refer to the difference between two versions of $B_N$ at the same face as ``B error''.

In order to solve for all changes to a cell's magnetic field, fluxes must be calculated not only in/out of the cell itself but also between all neighbors sharing one or more edges with the cell.
Therefore the MHD state of a cell with substep level $L_i$ can only be updated if substep level of all of its neighbors is at least $L_i$, otherwise the state update must wait for flux calculations between its relevant neighbors at substep level $L_i - 1$.
Here we assume a maximum difference of one between the substep levels of neighboring cells.
Table \ref{table:substepping} shows a step-by-step example of one simulation timestep divided into four substeps.

\begin{table}
\begin{tabular}{|c|c|c|}
\hline
Substep (\#) & Flux multiplier ($\delta t$) & Substep level ($L_i$) \\
\hline
1 & 1 & $2_* \leftrightarrow 2_*$ \\
 & - & $2_2 \Leftarrow 2_2$ \\
\hline
2 & 1 & $2_* \leftrightarrow 2_*$ \\
 & 1 & $1_* \leftrightarrow 2_*$ \\
 & 2 & $1_* \leftrightarrow 1_*$ \\
 & - & $2_* \Leftarrow 2_*$ \\
 & - & $1_1 \Leftarrow 1_1$ \\
\hline
3 & 1 & $2_* \leftrightarrow 2_*$ \\
 & - & $2_2 \Leftarrow 2_2$ \\
\hline
4 & 1 & $2_* \leftrightarrow 2_*$ \\
 & 1 & $1_* \leftrightarrow 2_*$ \\
 & 2 & $1_* \leftrightarrow 1_*$ \\
 & 2 & $0_* \leftrightarrow 1_*$ \\
 & 4 & $0_* \leftrightarrow 0_*$ \\
 & - & $2_* \Leftarrow 2_*$ \\
 & - & $1_* \Leftarrow 1_*$ \\
 & - & $0_0 \Leftarrow 0_0$ \\
\hline
\end{tabular}
\caption{
Step-by-step example of solving a timestep with $L_{max} = 2$, i.e. three substep levels and 4 substeps numbered sequentially from top to bottom in left column.
Middle column indicates the temporal multiplier of MHD fluxes in units of $\delta t$ or $-$ if it is not used.
In right side column, flux calculations between cells of corresponding substep level(s) are indicated by $\leftrightarrow$, while $\Leftarrow$ indicates that changes collected so far to cells' MHD state are applied.
Cells are indicated as $N_M$ where $N$ is cell's substep level and $M$ is maximum substep level of cell itself and its nearby neighbors, or $*$ for any value in range $0...L_{max}$.
During odd-numbered substeps, fluxes are only calculated between cells of shortest timestep, i.e. substep level $L_{max}$, with a flux multiplier of $1$.
Note that the MHD state is updated only in cells whose nearby neighbors also have substep level of $L_{max}$.
In substep 2 on the other hand, the MHD state of all cells of substep level $L_{max}$ is updated since fluxes have also been solved between cells of substep level $L_{max} - 1$.
In substep 2 fluxes are calculated between all cells with substep level $\geq L_{max} - 1$, with with flux multiplier of $2$.
In substep 4 all fluxes are solved and all cells are updated, after which a new timestep begins from substep 1.
}
\label{table:substepping}
\end{table}

\section{Cell-based mesh refinement}
\label{sec:amr}

Our model is implemented on top of DCCRG \cite{dccrg} and thus follows its AMR logic, i.e.~each cell can be refined independently of others into $2^3$ smaller cells and vice versa, as long as the logical size difference between neighboring cells is at most a factor of two.
The parallel computation functionality of DCCRG is not used here.

Top right panel in figure \ref{fig:grid} shows an example with mesh refinement of how one flux calculation (arrow) affects the face magnetic fields (black bars) of neighboring cells.
Below that we illustrate our approach to handling the missing fourth face indicated by a dashed red line that touches the edge indicated by a red circle.
Flux through the missing face is calculated as-if the large cell consisted of smaller cells with identical MHD state and affects 6 real faces.
Bottom right panel of figure \ref{fig:grid} shows all fluxes required for advancing the small middle cell's state by one step.
We note that due to the extended range of the effect of flux calculation in our approach, the largest timestep / smallest substep level allowed in each cell (CFL) is limited not only by the maximum wave speeds and physical size of the cell itself but also by those of its affected neighbors.

\section{Numerical tests}
\label{sec:tests}

The results presented here were obtained with version mhd\_v1 of \href{https://github.com/fmihpc/pamhd/releases/tag/mhd_v1}{github.com/fmihpc/pamhd} and are available at \href{https://doi.org/10.5281/zenodo.17183043}{doi:10.5281/zenodo.17183043} along with configuration files and post-processing scripts.
The AMR functionality of our model is provided by DCCRG \cite{dccrg}, and the generic simulation cell method described in \cite{honkonen15} has also simplified model development significantly.

We use the Rusanov ideal MHD solver, vacuum permeability $\mu_0 = 1$, adiabatic index $\gamma = 5/3$, dynamic timestep with CFL factor of $0.5$ and define $r = \sqrt{x^2+y^2}$.
In order to reduce initial errors in magnetic field to smaller values than produced during simulations, the face-centered fields in each cell are initialized using non-adaptive 7-point Gauss-Legendre integration in both dimensions.

For each test we run low and high resolution simulations with uniform meshes of $N^2$ and $(4N)^2$ cells respectively without substepping.
We also run simulations with substepping on statically refined meshes using different refinement criteria with effective lowest resolution of $N^2$ and highest resolution of $(4N)^2$.

\subsection{Alfvén wave}
\label{sec:alfven}

We use a circularly polarized traveling Alfvén wave with an angle of $30^{\circ}$ from $X$ axis as given in \cite{toth00}.
The simulation domain is periodic with $0 \leq x \leq 2/\sqrt{3}$ and $0 \leq y \leq 2$.
The initial condition is:
\begin{align*}
\rho = 10 p &= 1 \\
V_x &= -sin[\pi (\sqrt{3}x+y)]/20 \\
V_y &= -\sqrt{3} V_x \\
V_z = B_z &= cos[\pi(\sqrt{3}x+y)]/10 \\
B_x &= \sqrt{3}/2 + V_x \\
B_y &= 1/2 + V_y
\end{align*}
and the simulation is run from $t=0$ to $t=1$.
The wave propagates in $(-x, -y)$ direction.

\begin{figure}
\centering
\includegraphics[width = \textwidth]{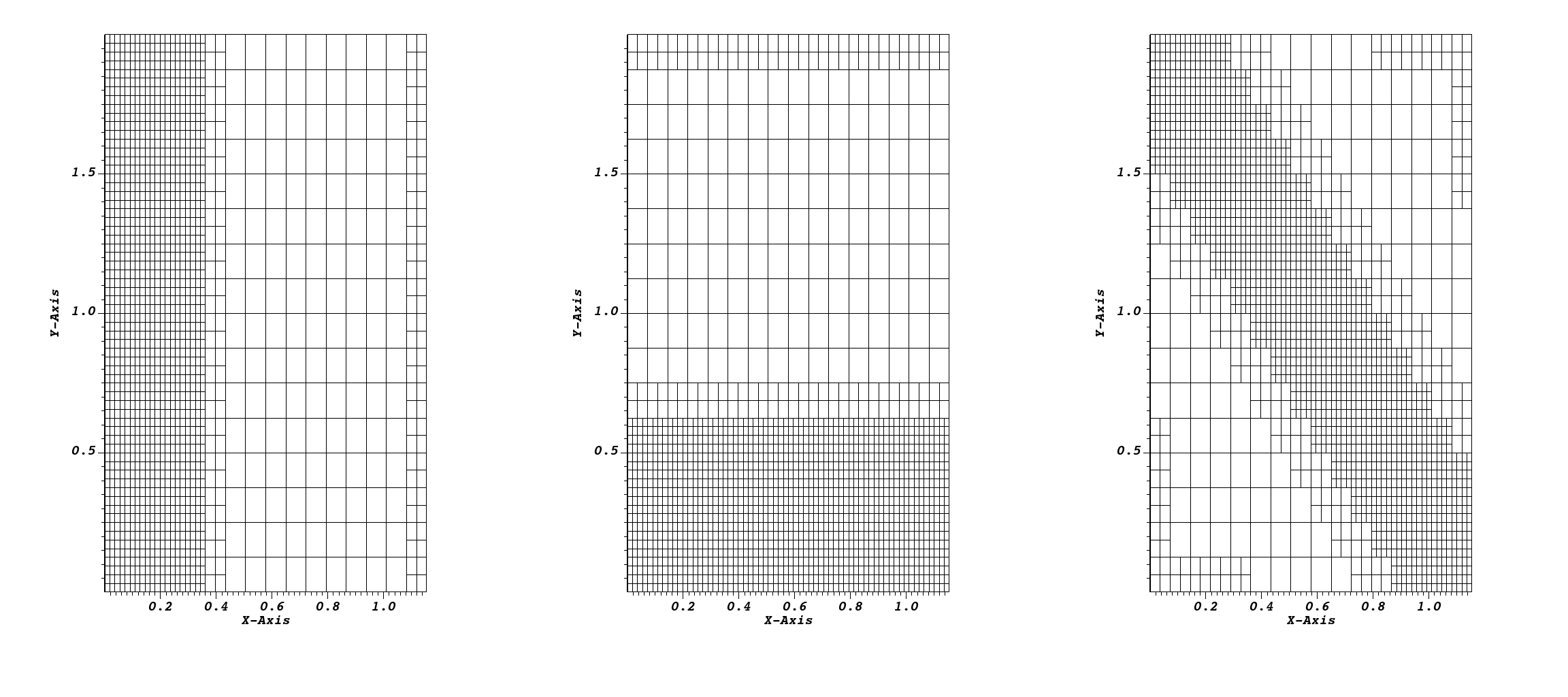}
\caption{
Refined meshes used in Alfvén wave test, runs from left to right: amr1, amr2 and amr3.
}
\label{fig:alfvengrid}
\end{figure}

We use $16^2$ cells in low resolution simulation and $64^2$ cells in high resolution.
We run 3 simulations with different refined meshes, shown in figure \ref{fig:alfvengrid}, times two simulations with substepping enabled and disabled:
Runs amr1 and sub\_amr1 use maximum resolution in the region $x < 2/\sqrt{27}$ and minimum resolution everywhere else,
runs amr2 and sub\_amr2 use maximum resolution in $y < 2/3$
and runs amr3 and sub\_amr3 use max.~resolution in $|\sqrt{3} x + y - 2| < 0.4$.
Divergence of magnetic field at $t=1$ is $\sim 10^{-14}$ in low resolution regions and $\sim 10^{-13}$ in high resolution regions in all runs (not shown).
B error is $\sim 10^{-14}$ and is mostly limited to cells with neighbors of different size.
The $64^2$ cell run calculates $\approx 1.5\cdot10^6$ MHD fluxes, runs amr1 and amr2 calculate $\approx 6.1\cdot10^5$ fluxes, runs sub\_amr1 and sub\_amr2 $\approx 5.6\cdot10^5$ fluxes, run amr3 $\approx 8.0\cdot10^5$ fluxes and run sub\_amr3 $\approx 7.6\cdot10^5$ fluxes.

\begin{figure}
\centering
\includegraphics[width = \textwidth]{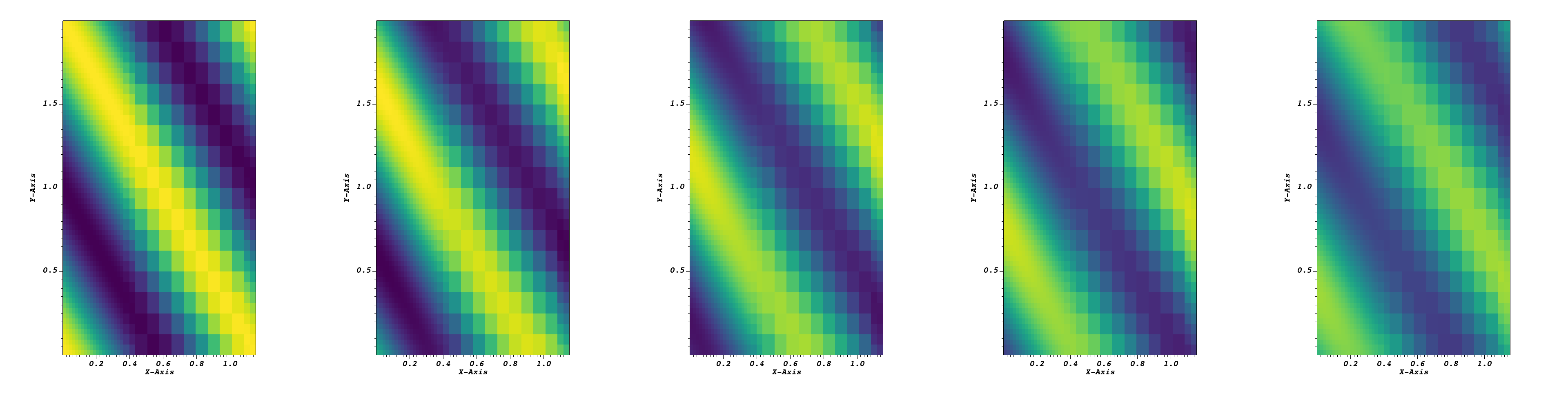}
\caption{
Snapshots of B$_z$ in run sub\_amr1 of circularly polarized traveling Alfvén wave test with color scale ranging from $-0.1$ to $0.1$.
Simulation time from left to right: $t=0.0$, $t=0.2$, $t=0.4$, $t=0.6$, $t=0.8$.
}
\label{fig:alfven}
\end{figure}

\begin{figure}
\centering
\begin{tabular}{cc}
\includegraphics[width=0.45\textwidth]{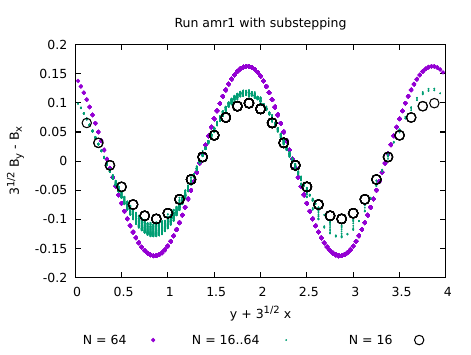} &
\includegraphics[width=0.45\textwidth]{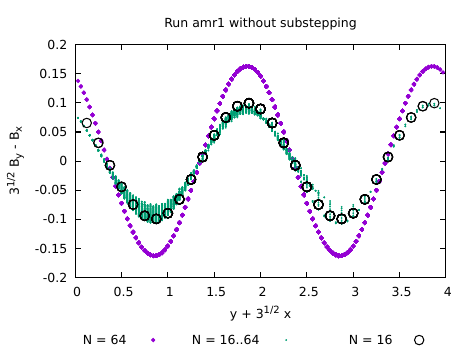} \\
\includegraphics[width=0.45\textwidth]{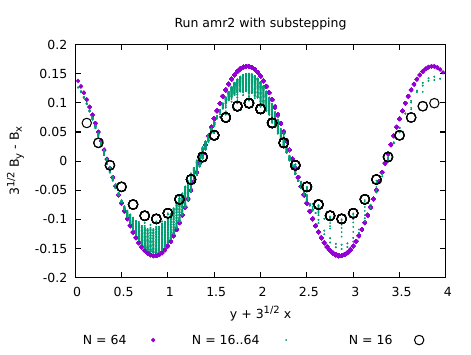} &
\includegraphics[width=0.45\textwidth]{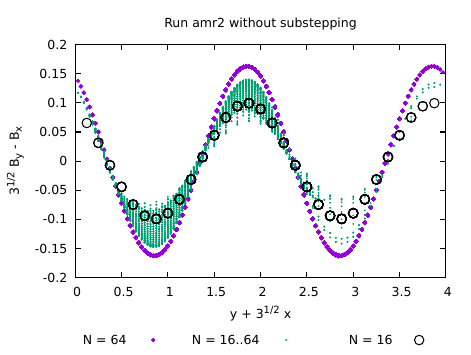} \\
\includegraphics[width=0.45\textwidth]{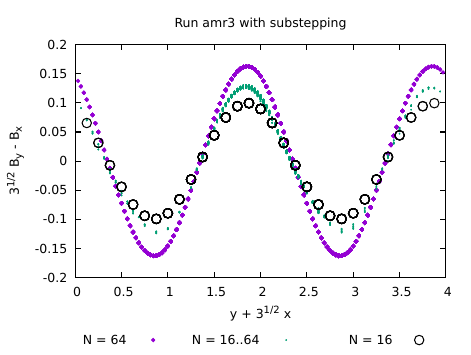} &
\includegraphics[width=0.45\textwidth]{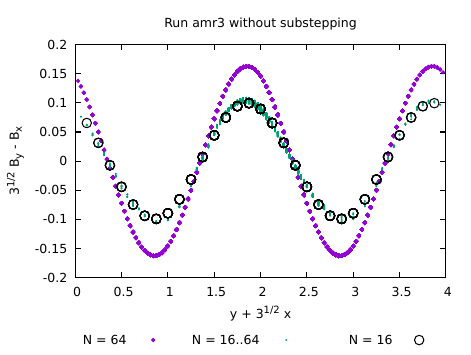} \\
\end{tabular}
\caption{
Circularly polarized traveling Alfvén wave test at $t=1$.
}
\label{fig:alfven2}
\end{figure}

Figure \ref{fig:alfven} shows snapshots of $B_z$ in run sub\_amr1.
Figure \ref{fig:alfven2} shows the effect of different meshes and substepping on the magnetic field perpendicular to the wave vector.
Specific behavior of the Alfvén wave in each run depends on the details of the refined mesh.
When substepping is enabled though, all simulations with refined mesh behave as one would expect, i.e.~their numerical diffusion is always between that of the low and high resolution simulations with uniform grid.
Additionally in our tests with refined mesh, numerical diffusion is smallest when resolution is uniform in the direction perpendicular to the wave vector.

\subsection{Blast wave}

In the blast wave test we use the initial condition given in \cite{londrillo00} but rotate the magnetic field similarly to the 2d shock tube test used in \cite{toth00}.
The simulation domain is periodic with $-0.5 \leq x,y \leq 0.5$, the initial condition is:
\begin{align*}
\rho &= 1 \\
p &=
\begin{cases}
r < 0.125: 100 \\
r \geq 0.125: 1
\end{cases}\\
V_x = V_y = V_z = B_z &= 0 \\
B_x &= 10 sin[tan^{-1}(2)] \\
B_y &= 10 cos[tan^{-1}(2)]
\end{align*}
and the simulation is run from $t=0$ to $t=0.02$.

\begin{figure}
\centering
\includegraphics[width = \textwidth]{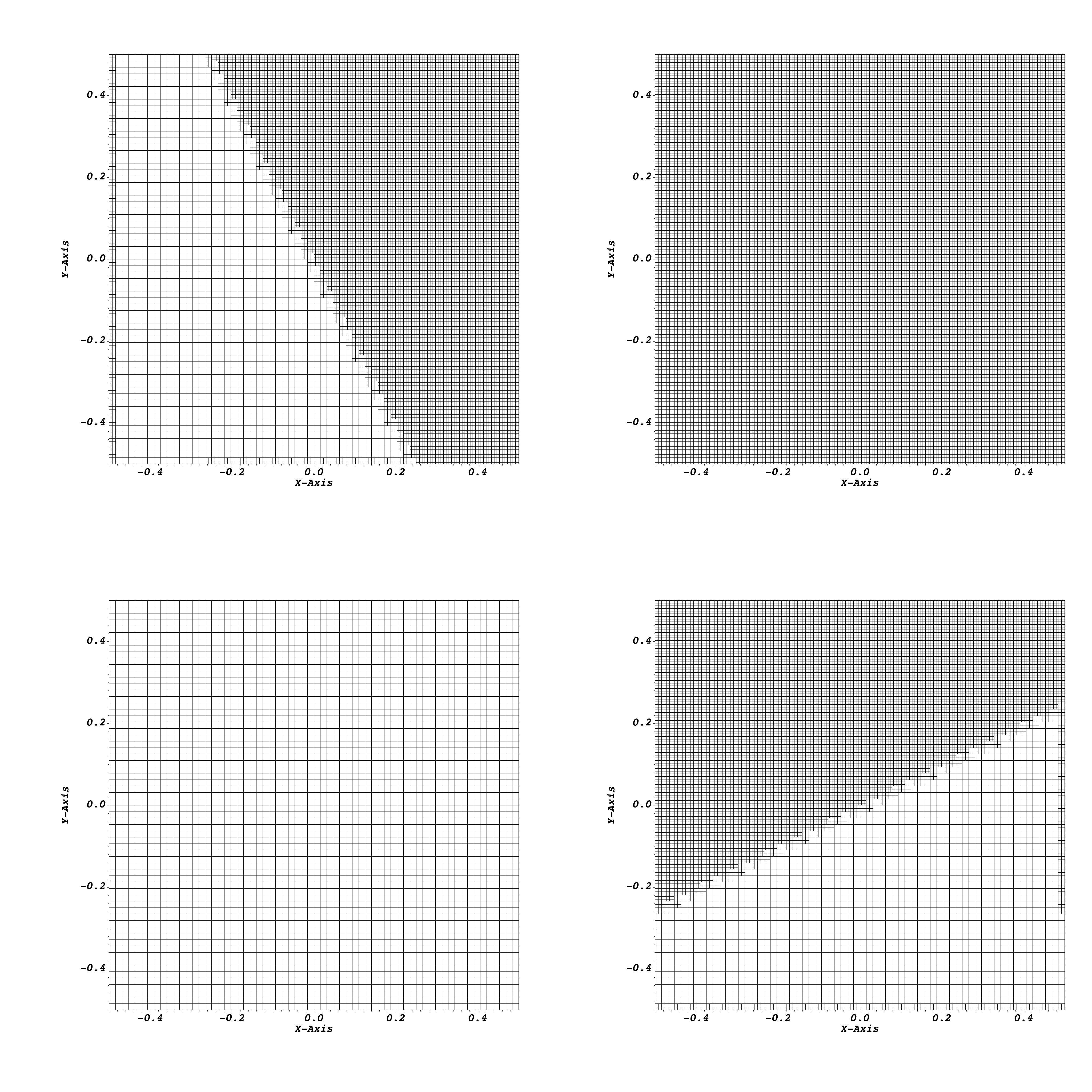}
\caption{
Meshes used in blast wave test, runs from left to right, top to bottom: amr1, high resolution, low resolution and amr2.
}
\label{fig:blastgrid}
\end{figure}

The low resolution simulation uses $64^2$ cells, high resolution $256^2$ cells.
We run two simulations with refined mesh and substepping shown in figure \ref{fig:blastgrid}: run amr1 has maximum resolution in the region $y + 2x > 0$ and minimum everywhere else while amr2 has maximum resolution in $x - 2y < 0$.
Divergence of magnetic field at $t=0.02$ is $\sim 10^{-12}$ in low resolution regions and $\sim 10^{-11}$ in high resolution regions in all runs (not shown).
B error is $\sim 10^{-13}$ and is mostly limited to cells with neighbors of different size.
The high resolution run solves $\approx 3.7 \cdot 10^7$ fluxes while runs amr1 and amr2 solve $\approx 1.9 \cdot 10^6$ fluxes.

\begin{figure}
\centering
\begin{tabular}{cc}
\includegraphics[width=0.45\textwidth]{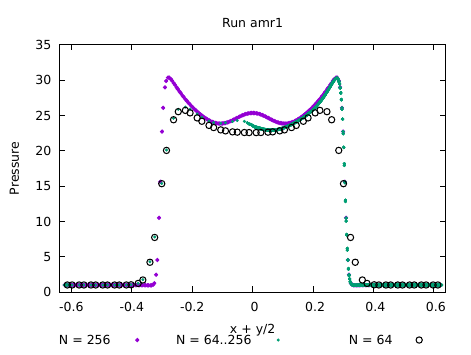} &
\includegraphics[width=0.45\textwidth]{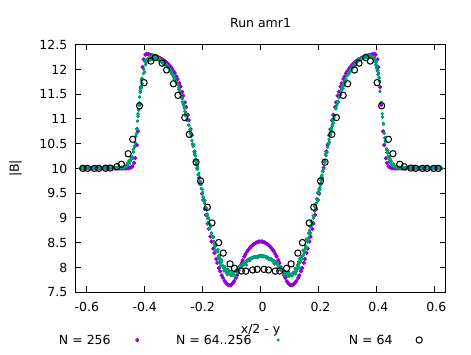} \\
\includegraphics[width=0.45\textwidth]{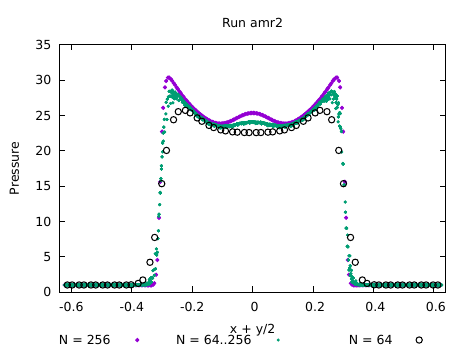} &
\includegraphics[width=0.45\textwidth]{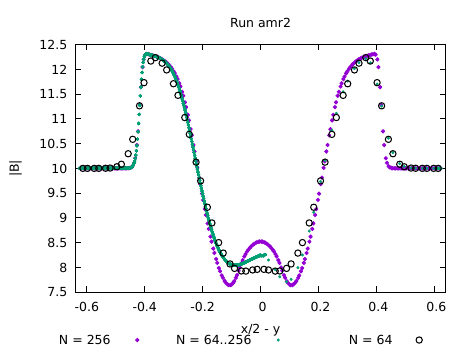} \\
\end{tabular}
\caption{
Blast wave test at $t=0.02$.
}
\label{fig:blast1}
\end{figure}

\begin{figure}
\centering
\includegraphics[width = \textwidth]{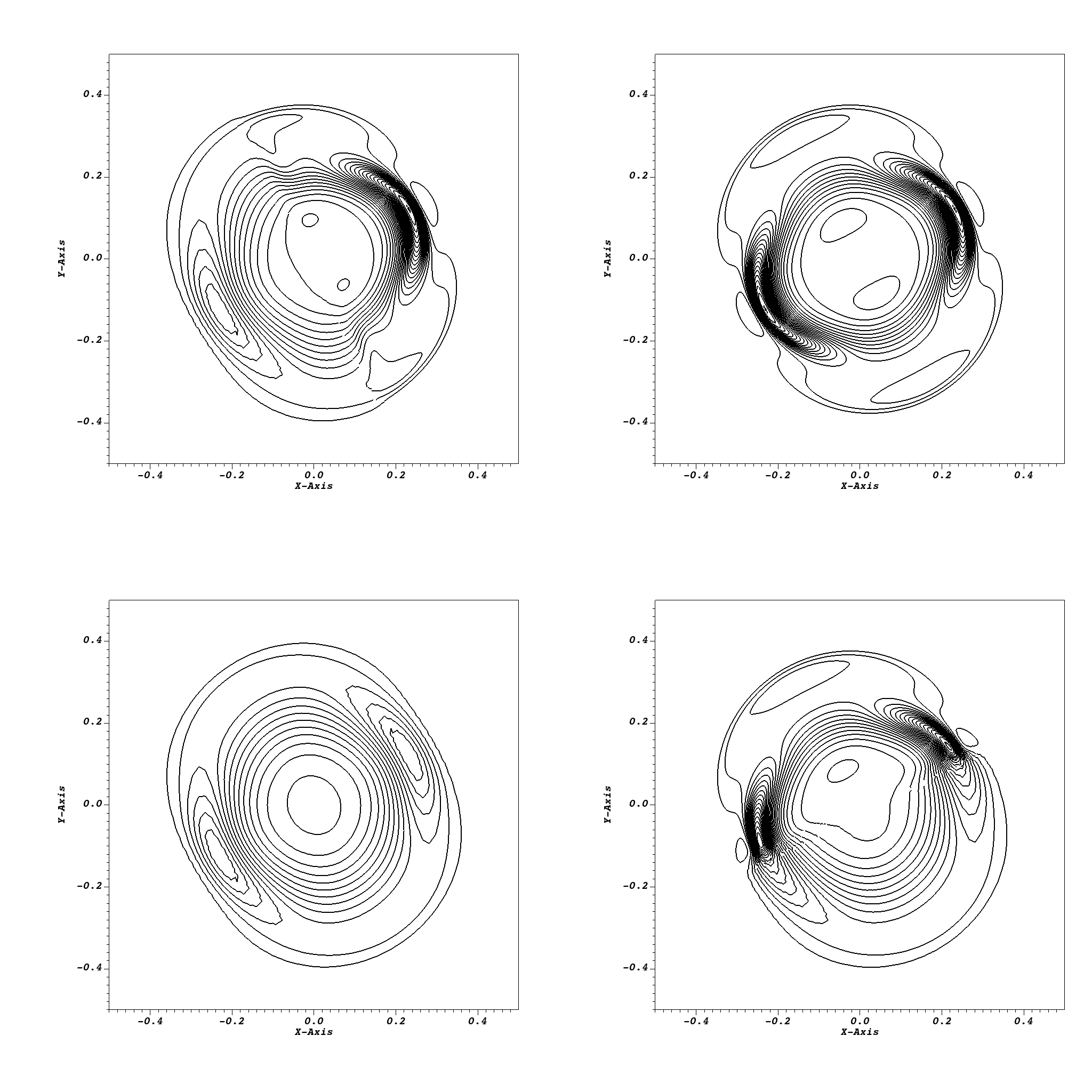}
\caption{
Contours of density in blast wave test with 30 levels from 0.3 to 2.5 at $t=0.02$.
Different grids from left to right, top to bottom:
amr1, high resolution, low resolution, amr2.
}
\label{fig:blast2}
\end{figure}

Figure \ref{fig:blast1} shows 1d cuts of pressure and magnetic field magnitude through middle of the domain parallel and perpendicular to initial magnetic field and figure \ref{fig:blast2} shows contours of density.
The model behaves as one would expect.
For example the leading edge of the blast wave in $+x,+y$ direction looks identical in both the high-resolution run and amr1.
Similarly, the leading edge in $-x,+y$ direction looks identical in high resolution and amr2 runs.
The non-coordinate -aligned mesh refinement boundary does not seem to affect the solution except for higher numerical diffusion on the side of lower resolution.

\subsection{Orszag-Tang vortex}

In the Orszag-Tang vortex test we use the initial condition given in \cite{toth00}.
The simulation domain is periodic with $0 \leq x,y \leq 2 \pi$, the initial condition is:
\begin{align*}
\rho &= \gamma^2 \\
p &= \gamma \\
V_x = B_x &= -sin(y) \\
V_y &= sin(x) \\
B_y &= sin(2 x) \\
V_z = B_z &= 0
\end{align*}
and the simulation is run from $t=0$ to $t=3.14$.

\begin{figure}
\centering
\includegraphics[width = \textwidth]{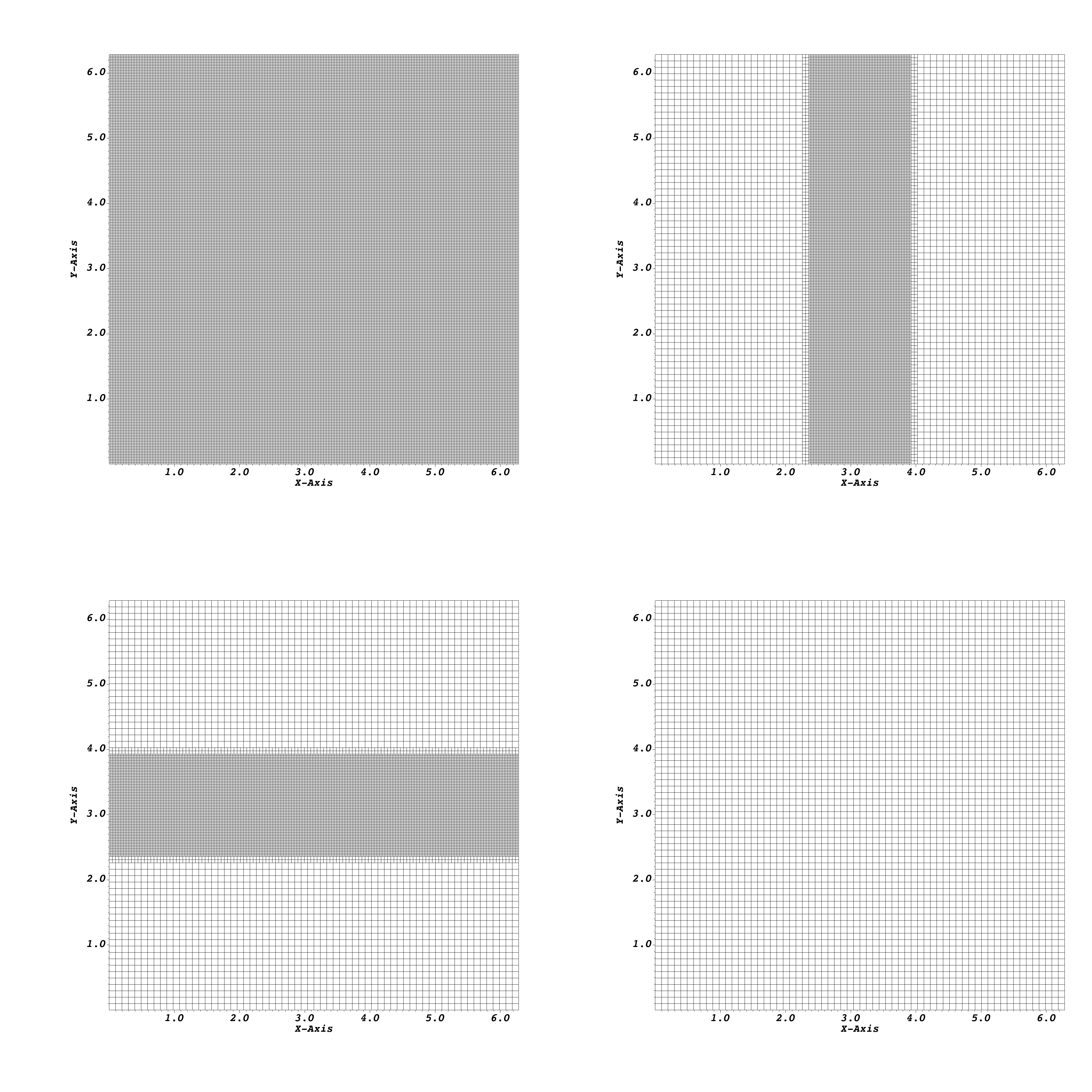}
\caption{
Meshes used in Orszag-Tang vortex test, runs from left to right, top to bottom: high resolution, amr1, amr2, low resolution.
}
\label{fig:otgrid}
\end{figure}

We use $64^2$ cells in low resolution simulation and $256^2$ in high resolution.
We run two simulations with refined mesh and substepping shown in figure \ref{fig:otgrid}:
run amr1 uses maximum resolution in the region $|x - \pi| < \pi/4$ and minimum everywhere else, while run amr2 uses maximum resolution in $|y - \pi| < \pi/4$.
Divergence of magnetic field at $t=3.14$ is $\sim 10^{-14}$ in low resolution regions and $\sim 10^{-13}$ in high resolution regions in all runs (not shown).
B error is $\sim 10^{-14}$ and is mostly limited to cells with neighbors of different size.
The high resolution simulation calculates $\approx 1.2 \cdot 10^8$ fluxes, run amr1 $\approx 2.9 \cdot 10^7$ fluxes and run amr2 $\approx 3.0 \cdot 10^7$ fluxes.

\begin{figure}
\centering
\begin{tabular}{cc}
\includegraphics[width=0.45\textwidth]{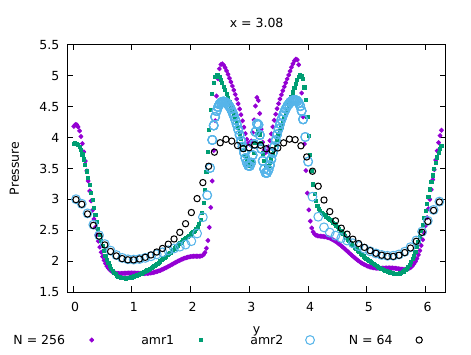} &
\includegraphics[width=0.45\textwidth]{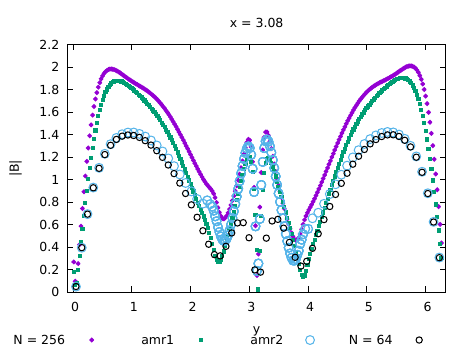} \\
\includegraphics[width=0.45\textwidth]{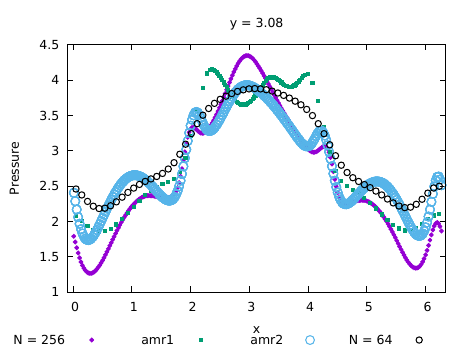} &
\includegraphics[width=0.45\textwidth]{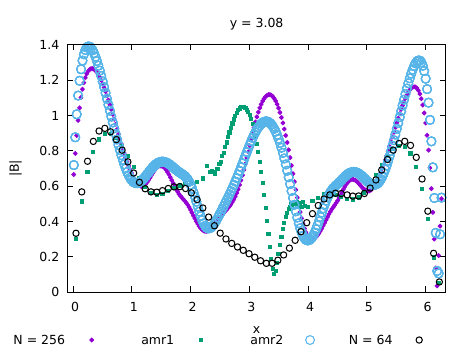} \\
\end{tabular}
\caption{
Orszag-Tang vortex test at $t=3.14$.
}
\label{fig:ot1}
\end{figure}

\begin{figure}
\centering
\includegraphics[width = \textwidth]{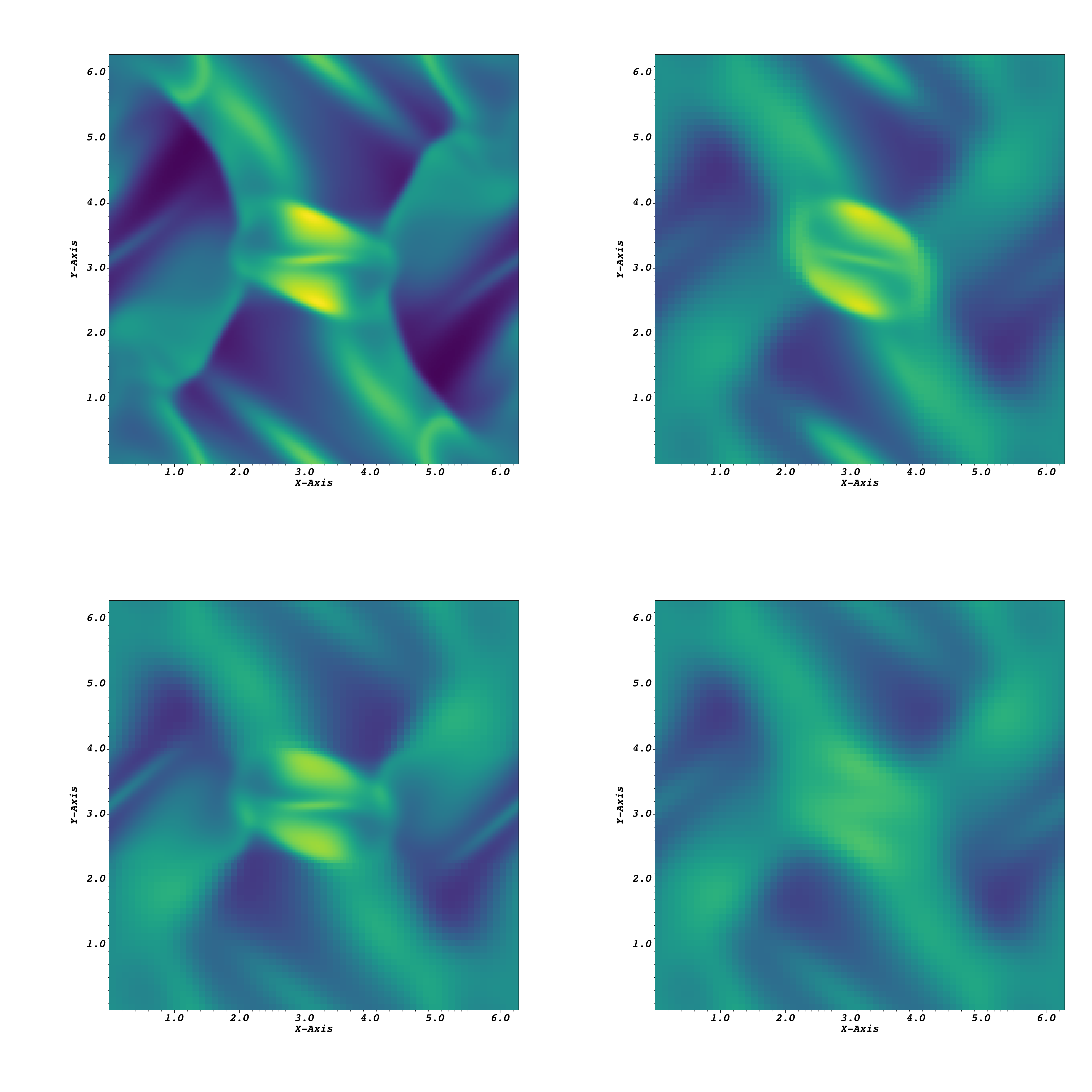}
\caption{
Pressure in Orszag-Tang vortex at $t=0.295$ with color scale ranging from 0.6 to 5.3.
Different grids from left to right, top to bottom: uniform high resolution, amr1, amr2 and low resolution.
}
\label{fig:ot2}
\end{figure}

Figure \ref{fig:ot1} shows 1d cuts of pressure and magnetic field magnitude through middle of the domain along coordinate directions and figure \ref{fig:ot2} shows 2d snapshots of pressure.
In run amr2 the solution resembles the uniform high resolution result, while run amr1 exhibits over-rotation in the middle.

\subsection{Magnetic rotor}

In the magnetic rotor test we use the setup of the second rotor test given in \cite{toth00}, shifted to the origin.
The simulation domain is periodic with $-0.5 \leq x,y \leq 0.5$, the initial condition is:
\begin{align*}
\rho &= f(1,1+600(0.115-r),10) \\
p &= 0.5 \\
V_x &= -100 y f(0,r,(0.115-r)/0.15) \\
V_y &= 100 x f(0,r,(0.115-r)/0.15) \\
V_z = B_y = B_z &= 0 \\
B_x &= 1.25/\sqrt{\pi}
\end{align*}
where
\begin{align*}
f(a,b,c) &= 
\begin{cases}
a > b: a \\
c < b: c \\
a \leq b \leq c: b
\end{cases}
\end{align*}
and the simulation is run from $t=0$ to $t=0.295$.

\begin{figure}
\centering
\includegraphics[width = \textwidth]{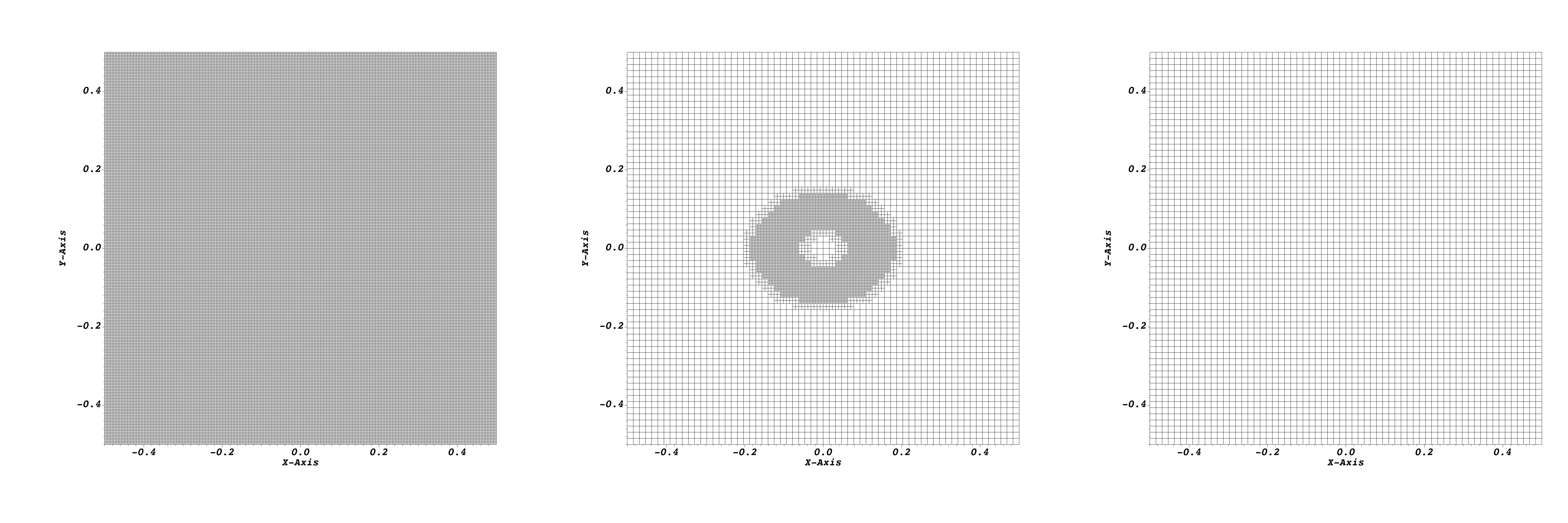}
\caption{
Meshes used in magnetic rotor test, runs from left to right: high resolution, amr1, low resolution.
}
\label{fig:rotorgrid}
\end{figure}

The low resolution simulation uses $64^2$ cells, high resolution $256^2$ cells and refined mesh simulation amr1 uses substepping and maximum resolution in the region $0.002 \leq 0.6x^2 + y^2 \leq 0.02$, shown in figure \ref{fig:rotorgrid}.
Divergence of magnetic field at $t=0.295$ is $\sim 10^{-12}$ in low resolution regions and $\sim 10^{-13}$ in high resolution regions in all runs (not shown).
B error is $\sim 10^{-14}$ and is mostly limited to cells with neighbors of different size.
The high resolution simulation calculates $4.8 \cdot 10^7$ fluxes while amr1 calculates $4.0 \cdot 10^6$ fluxes.

\begin{figure}
\centering
\begin{tabular}{cc}
\includegraphics[width=0.45\textwidth]{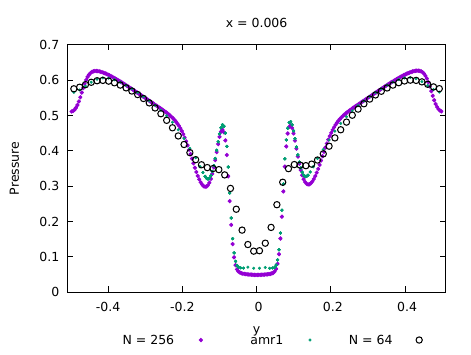} &
\includegraphics[width=0.45\textwidth]{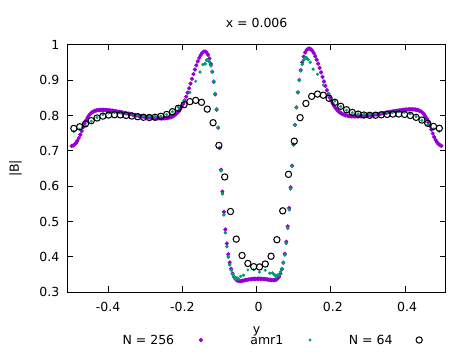} \\
\includegraphics[width=0.45\textwidth]{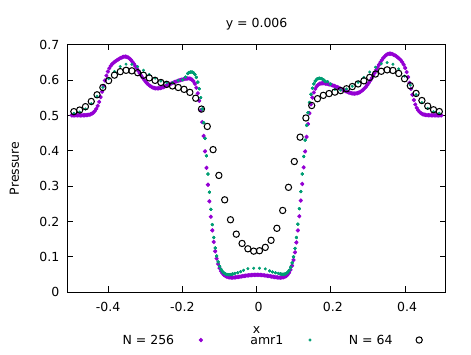} &
\includegraphics[width=0.45\textwidth]{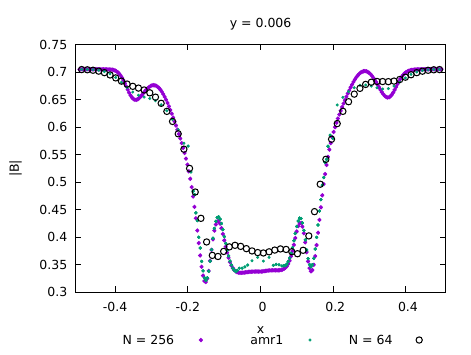} \\
\end{tabular}
\caption{
Magnetic rotor test at $t=0.295$
}
\label{fig:rotor1}
\end{figure}

\begin{figure}
\centering
\includegraphics[width = \textwidth]{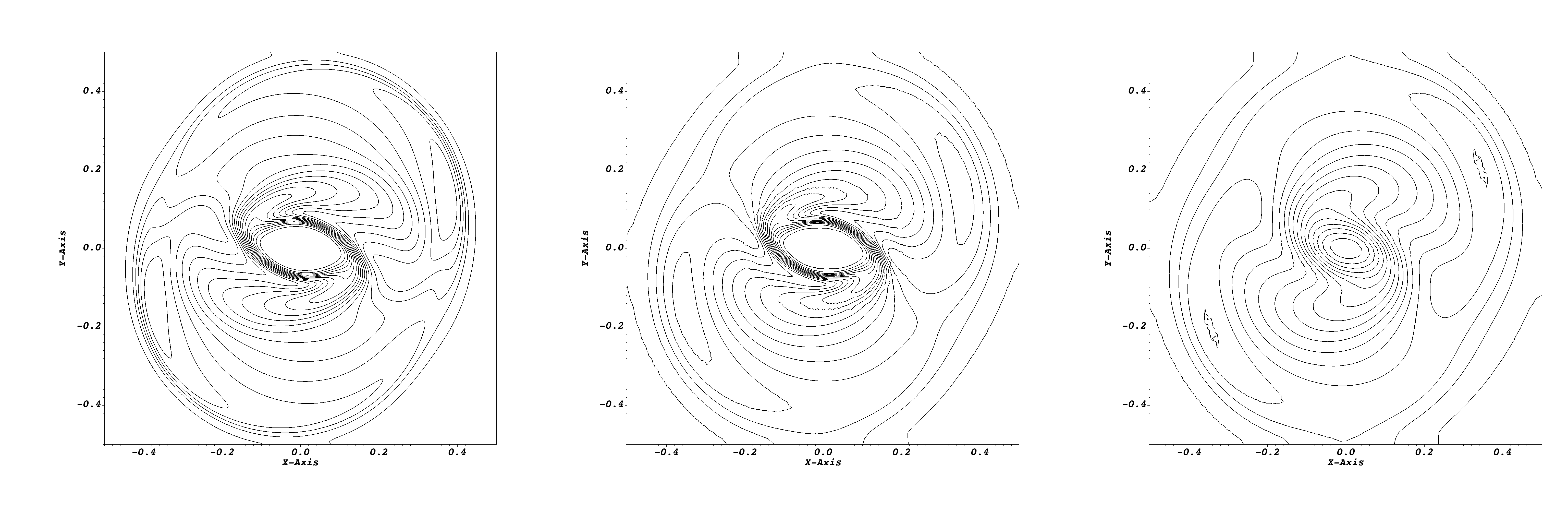}
\caption{
Contour of pressure in magnetic rotor test using 20 levels from 0.039 to 0.72 at $t=0.295$.
Different grids from left to right: high resolution, amr1, low resolution.
}
\label{fig:rotor2}
\end{figure}

Figure \ref{fig:rotor1} shows 1d cuts of pressure and magnetic field magnitude through middle of the domain along coordinate directions and figure \ref{fig:rotor2} shows 2d contours of pressure.
The solution calculated using refined mesh resembles that of the high resolution mesh, with higher numeric diffusion further away from the high resolution region.

\section{Discussion}
\label{sec:discussion}
As demonstrated in section \ref{sec:tests}, our approach enables a flexible tradeoff between physical accuracy of results and the computational resources required for a simulation.
Numerical diffusion can be minimized by using high resolution in region(s) of interest and temporal substepping where longer timesteps are allowed by the CFL condition.
As also shown in section \ref{sec:tests}, the details of mesh refinement can have a large effect on the result but we leave the investigation of optimal mesh refinement strategies as well as run-time AMR for future work.
Run-time AMR based on gradients of various plasma parameters has been successfully employed e.g.~by the GUMICS model for Earth's magnetosphere \cite{gumics} and for the MHD blast wave test in \cite{dccrg}.

Our staggered field solver should be applicable also to other approaches to AMR, such as block-based and patch-based refinement, as long as cells of the same logical size do not overlap and cells of different logical size are properly nested, i.e.~parent cells completely overlap their children, etc.
Some restrictions in the algorithms presented here may not be necessary but are used in order to simplify our implementation.
For example it should be possible to have cells refine into $N \times M \times L$ children where $N,M,L \geq 2$ instead of $N=M=L=2$.
Similarly, it might be possible for a cell to be solved an arbitrary number of times instead of $2^{L_i}$ times during one timestep.

\section{Conclusions}
\label{sec:conclusions}

We develop a staggered mesh magnetic field solver based on MHD fluxes that supports cell-based adaptive mesh refinement and temporal substepping.
Several standard numerical tests in two dimensions indicate that our approach enables a flexible tradeoff between computational resources and physical accuracy.
This is especially useful for multiscale systems, such as planetary magnetospheres, where the entire system should be modeled in order to properly capture its behavior, but high resolution can be concentrated in specific region(s) of interest.

\bibliographystyle{abbrv}
\bibliography{references.bib}

@article{marti15,
       author = {{Mart{\'\i}}, Jos{\'e} Mar{\'\i}a and {M{\"u}ller}, Ewald},
        title = "{Grid-based Methods in Relativistic Hydrodynamics and Magnetohydrodynamics}",
      journal = {Living Reviews in Computational Astrophysics},
         year = 2015,
        month = dec,
       volume = {1},
       number = {1},
          eid = {3},
        pages = {3},
          doi = {10.1007/lrca-2015-3}
}

@article{mignone21,
       author = {{Mignone}, A. and {Del Zanna}, L.},
        title = "{Systematic construction of upwind constrained transport schemes for MHD}",
      journal = {Journal of Computational Physics},
         year = 2021,
        month = jan,
       volume = {424},
          eid = {109748},
        pages = {109748},
          doi = {10.1016/j.jcp.2020.109748},
archivePrefix = {arXiv},
       eprint = {2004.10542},
 primaryClass = {physics.comp-ph}
}

@article{ryu95,
       author = {{Ryu}, Dongsu and {Jones}, T.~W. and {Frank}, Adam},
        title = "{Numerical Magnetohydrodynamics in Astrophysics: Algorithm and Tests for Multidimensional Flow}",
      journal = {Astrophysical Journal},
         year = 1995,
        month = oct,
       volume = {452},
        pages = {785},
          doi = {10.1086/176347},
archivePrefix = {arXiv},
       eprint = {astro-ph/9505073},
 primaryClass = {astro-ph}
}

@article{powell99,
       author = {{Powell}, Kenneth G. and {Roe}, Philip L. and {Linde}, Timur J. and {Gombosi}, Tamas I. and {De Zeeuw}, Darren L.},
        title = "{A Solution-Adaptive Upwind Scheme for Ideal Magnetohydrodynamics}",
      journal = {Journal of Computational Physics},
         year = 1999,
        month = sep,
       volume = {154},
       number = {2},
        pages = {284-309},
          doi = {10.1006/jcph.1999.6299}
}

@book{leveq2,
  author    = {LeVeque, Randall J.},
  title     = {Finite-Volume Methods for Hyperbolic Problems},
  publisher = {Cambridge University Press},
  year      = {2012},
        doi = {10.1017/CBO9780511791253},
        isbn= {9780511791253}
}

@article{athena20,
       author = {{Stone}, James M. and {Tomida}, Kengo and {White}, Christopher J. and {Felker}, Kyle G.},
        title = "{The Athena++ Adaptive Mesh Refinement Framework: Design and Magnetohydrodynamic Solvers}",
      journal = {The Astrophysical Journal Supplement Series},
         year = 2020,
        month = jul,
       volume = {249},
       number = {1},
          eid = {4},
        pages = {4},
          doi = {10.3847/1538-4365/ab929b},
archivePrefix = {arXiv},
       eprint = {2005.06651},
 primaryClass = {astro-ph.IM}
 }

@article{pluto12,
       author = {{Mignone}, A. and {Zanni}, C. and {Tzeferacos}, P. and {van Straalen}, B. and {Colella}, P. and {Bodo}, G.},
        title = "{The PLUTO Code for Adaptive Mesh Computations in Astrophysical Fluid Dynamics}",
      journal = {The Astrophysical Journal Supplement Series},
         year = 2012,
        month = jan,
       volume = {198},
       number = {1},
          eid = {7},
        pages = {7},
          doi = {10.1088/0067-0049/198/1/7},
archivePrefix = {arXiv},
       eprint = {1110.0740},
 primaryClass = {astro-ph.HE}
}

@article{flash00,
       author = {{Fryxell}, B. and {Olson}, K. and {Ricker}, P. and {Timmes}, F.~X. and {Zingale}, M. and {Lamb}, D.~Q. and {MacNeice}, P. and {Rosner}, R. and {Truran}, J.~W. and {Tufo}, H.},
        title = "{FLASH: An Adaptive Mesh Hydrodynamics Code for Modeling Astrophysical Thermonuclear Flashes}",
      journal = {The Astrophysical Journal Supplement Series},
         year = 2000,
        month = nov,
       volume = {131},
       number = {1},
        pages = {273-334},
          doi = {10.1086/317361}
}

@article{enzo14,
       author = {{Bryan}, Greg L. and {Norman}, Michael L. and {O'Shea}, Brian W. and {Abel}, Tom and {Wise}, John H. and {Turk}, Matthew J. and {Reynolds}, Daniel R. and {Collins}, David C. and {Wang}, Peng and {Skillman}, Samuel W. and {Smith}, Britton and {Harkness}, Robert P. and {Bordner}, James and {Kim}, Ji-hoon and {Kuhlen}, Michael and {Xu}, Hao and {Goldbaum}, Nathan and {Hummels}, Cameron and {Kritsuk}, Alexei G. and {Tasker}, Elizabeth and {Skory}, Stephen and {Simpson}, Christine M. and {Hahn}, Oliver and {Oishi}, Jeffrey S. and {So}, Geoffrey C. and {Zhao}, Fen and {Cen}, Renyue and {Li}, Yuan and {Enzo Collaboration}},
        title = "{ENZO: An Adaptive Mesh Refinement Code for Astrophysics}",
      journal = {The Astrophysical Journal Supplement Series},
         year = 2014,
        month = apr,
       volume = {211},
       number = {2},
          eid = {19},
        pages = {19},
          doi = {10.1088/0067-0049/211/2/19},
archivePrefix = {arXiv},
       eprint = {1307.2265},
 primaryClass = {astro-ph.IM}
}

@ARTICLE{amrvac18,
       author = {{Xia}, C. and {Teunissen}, J. and {El Mellah}, I. and {Chan{\'e}}, E. and {Keppens}, R.},
        title = "{MPI-AMRVAC 2.0 for Solar and Astrophysical Applications}",
      journal = {The Astrophysical Journal Supplement Series},
         year = 2018,
        month = feb,
       volume = {234},
       number = {2},
          eid = {30},
        pages = {30},
          doi = {10.3847/1538-4365/aaa6c8},
archivePrefix = {arXiv},
       eprint = {1710.06140},
 primaryClass = {astro-ph.SR}
}

@article{harlow65,
author = {{Harlow}, Francis H. and {Welch}, J. Eddie},
title = "{Numerical Calculation of Time-Dependent Viscous Incompressible Flow of Fluid with Free Surface}",
journal = {Physics of Fluids},
year = 1965,
month = dec,
volume = {8},
number = {12},
pages = {2182-2189},
doi = {10.1063/1.1761178}
}

@article{yee66,
author = {{Yee}, Kane},
title = "{Numerical solution of initial boundary value problems involving maxwell's equations in isotropic media}",
journal = {IEEE Transactions on Antennas and Propagation},
year = 1966,
month = may,
volume = {14},
number = {3},
pages = {302-307},
doi = {10.1109/TAP.1966.1138693},
}

@article{evans88,
author = {{Evans}, Charles R. and {Hawley}, John F.},
title = "{Simulation of Magnetohydrodynamic Flows: A Constrained Transport Model}",
journal = {The Astrophysical Journal},
year = 1988,
month = sep,
volume = {332},
pages = {659},
doi = {10.1086/166684}
}

@article{balsarakim04,
author = {{Balsara}, Dinshaw S. and {Kim}, Jongsoo},
title = "{A Comparison between Divergence-Cleaning and Staggered-Mesh Formulations for Numerical Magnetohydrodynamics}",
journal = {The Astrophysical Journal},
year = 2004,
month = feb,
volume = {602},
number = {2},
pages = {1079-1090},
doi = {10.1086/381051},
archivePrefix = {arXiv},
eprint = {astro-ph/0310728},
primaryClass = {astro-ph},
}

@book{lipatov02,
author = {{Lipatov}, Alexander S.},
title = "{The hybrid multiscale simulation technology: an introduction with application to astrophysical and laboratory plasmas}",
publisher = {Springer-Verlag},
address = {Berlin, Heidelberg},
year = 2002,
}

@article{toth12,
author = {{T{\'o}th}, G{\'a}bor and {van der Holst}, Bart and {Sokolov}, Igor V. and {De Zeeuw}, Darren L. and {Gombosi}, Tamas I. and {Fang}, Fang and {Manchester}, Ward B. and {Meng}, Xing and {Najib}, Dalal and {Powell}, Kenneth G. and {Stout}, Quentin F. and {Glocer}, Alex and {Ma}, Ying-Juan and {Opher}, Merav},
title = "{Adaptive numerical algorithms in space weather modeling}",
journal = {Journal of Computational Physics},
year = 2012,
month = feb,
volume = {231},
number = {3},
pages = {870-903},
doi = {10.1016/j.jcp.2011.02.006},
}

@Article{honkonen15,
AUTHOR = {Honkonen, I.},
TITLE = {A generic simulation cell method for developing extensible, efficient and readable parallel computational models},
JOURNAL = {Geoscientific Model Development},
VOLUME = {8},
YEAR = {2015},
NUMBER = {3},
PAGES = {473--483},
URL = {https://doi.org/10.5194/gmd-8-473-2015},
DOI = {10.5194/gmd-8-473-2015}
}

@ARTICLE{cfl,
       author = {{Courant}, R. and {Friedrichs}, K. and {Lewy}, H.},
        title = "{On the Partial Difference Equations of Mathematical Physics}",
      journal = {IBM Journal of Research and Development},
         year = 1967,
        month = mar,
       volume = {11},
        pages = {215-234},
          doi = {10.1147/rd.112.0215},
}

@article{brecht81,
author = {Brecht, S. H. and Lyon, J. and Fedder, J. A. and Hain, K.},
title = "{A simulation study of east-west IMF effects on the magnetosphere}",
journal = "{Geophysical Research Letters}",
volume = {8},
number = {4},
pages = {397-400},
doi = {10.1029/GL008i004p00397},
url = {https://doi.org/10.1029/GL008i004p00397},
eprint = {https://agupubs.onlinelibrary.wiley.com/doi/pdf/10.1029/GL008i004p00397},
year = {1981}
}

@article{tanaka94,
title = {Finite Volume TVD Scheme on an Unstructured Grid System for Three-Dimensional MHD Simulation of Inhomogeneous Systems Including Strong Background Potential Fields},
journal = {Journal of Computational Physics},
volume = {111},
number = {2},
pages = {381-389},
year = {1994},
issn = {0021-9991},
doi = {10.1006/jcph.1994.1071},
url = {https://doi.org/10.1006/jcph.1994.1071},
author = {T. Tanaka},
}

@article{balsara99,
title = {A Staggered Mesh Algorithm Using High Order Godunov Fluxes to Ensure Solenoidal Magnetic Fields in Magnetohydrodynamic Simulations},
journal = {Journal of Computational Physics},
volume = {149},
number = {2},
pages = {270-292},
year = {1999},
issn = {0021-9991},
doi = {10.1006/jcph.1998.6153},
url = {https://doi.org/10.1006/jcph.1998.6153},
author = {Dinshaw S Balsara and Daniel S Spicer},
}

@article{londrillo00,
doi = {10.1086/308344},
url = {https://dx.doi.org/10.1086/308344},
year = {2000},
month = {feb},
volume = {530},
number = {1},
pages = {508},
author = {Londrillo, P. and Del Zanna, L.},
title = {High-Order Upwind Schemes for
Multidimensional Magnetohydrodynamics},
journal = {The Astrophysical Journal},
}

@article{toth00,
title = "{The div(B)=0 Constraint in Shock-Capturing Magnetohydrodynamics Codes}",
journal = {Journal of Computational Physics},
volume = {161},
number = {2},
pages = {605-652},
year = {2000},
issn = {0021-9991},
doi = {10.1006/jcph.2000.6519},
url = {https://doi.org/10.1006/jcph.2000.6519},
author = {Gábor Tóth}
}

@article{balsara01,
title = {Divergence-Free Adaptive Mesh Refinement for Magnetohydrodynamics},
journal = {Journal of Computational Physics},
volume = {174},
number = {2},
pages = {614-648},
year = {2001},
issn = {0021-9991},
doi = {10.1006/jcph.2001.6917},
url = {https://doi.org/10.1006/jcph.2001.6917},
author = {Dinshaw S. Balsara},
}

@article{toth02,
title = {Divergence- and Curl-Preserving Prolongation and Restriction Formulas},
journal = {Journal of Computational Physics},
volume = {180},
number = {2},
pages = {736-750},
year = {2002},
issn = {0021-9991},
doi = {10.1006/jcph.2002.7120},
url = {https://doi.org/10.1006/jcph.2002.7120},
author = {G. Tóth and P.L. Roe},
}

@article{balsara04,
doi = {10.1086/381377},
url = {https://dx.doi.org/10.1086/381377},
year = {2004},
month = {mar},
publisher = {American Astronomical Society},
volume = {151},
number = {1},
pages = {149},
author = {Balsara, Dinshaw S.},
title = {Second-Order-accurate Schemes for Magnetohydrodynamics with Divergence-free Reconstruction},
journal = {The Astrophysical Journal Supplement Series}
}

@article{gumics,
title = {The GUMICS-4 global MHD magnetosphere–ionosphere coupling simulation},
journal = {Journal of Atmospheric and Solar-Terrestrial Physics},
volume = {80},
pages = {48-59},
year = {2012},
issn = {1364-6826},
doi = {10.1016/j.jastp.2012.03.006},
url = {https://doi.org/10.1016/j.jastp.2012.03.006},
author = {P. Janhunen and M. Palmroth and T. Laitinen and I. Honkonen and L. Juusola and G. Facskó and T.I. Pulkkinen},
}

@article{dccrg,
title = {Parallel grid library for rapid and flexible simulation development},
journal = {Computer Physics Communications},
volume = {184},
number = {4},
pages = {1297-1309},
year = {2013},
issn = {0010-4655},
doi = {10.1016/j.cpc.2012.12.017},
url = {https://www.sciencedirect.com/science/article/pii/S0010465512004237},
author = {I. Honkonen and S. {von Alfthan} and A. Sandroos and P. Janhunen and M. Palmroth}
}
\end{document}